\documentclass[a4paper,10pt]{jpconf}
\usepackage{amsmath,amssymb,amsfonts}
\usepackage{graphicx}
\usepackage[mathscr]{eucal}
\usepackage[lofdepth,lotdepth]{subfig}
\usepackage{color}

\def\ben{\begin{equation}}
\def\een{\end{equation}}
\def\bea{\begin{eqnarray}}
\def\eea{\end{eqnarray}}

\begin{document}
\title{Curved geometry and Graphs\footnote{Based on a talk given at Loops '11, Madrid, on 24 May 2011.}}

\author{Francesco Caravelli}

\address{Perimeter Institute for Theoretical Physics, \\
Waterloo, Ontario N2L 2Y5
Canada, \\and\\
 University of Waterloo, Waterloo, Ontario N2L 3G1, Canada,\\
 and\\
 Max Planck Institute for Gravitational Physics, Albert Einstein Institute,\\
Am M\"uhlenberg 1, Golm, D-14476 Golm, Germany}

\ead{fcaravelli@perimeterinstitute.ca}


\begin{abstract}
Quantum Graphity is an approach to quantum gravity based on a background independent formulation of 
condensed matter systems on graphs. We summarize  recent results obtained on the notion of emergent geometry from
the point of view of a particle hopping on the graph. We discuss the role of connectivity in emergent Lorentzian 
perturbations in a curved background and the Bose--Hubbard (BH) model defined on graphs with particular symmetries.
\end{abstract}


\section{Introduction}
The gravitational interaction may be an effective 
description of an underlying 
theory which does not suffer from the well-known problems plaguing Einstein's gravity, as for instance perturbative 
non-renormalizability \footnote{However, there is still hope that gravity is non-perturbativly renormalizable. 
This is the starting point of 
Asymptotic Safety and Loop Quantum Gravity\cite{qg,as}.}. These ideas are the starting point of Quantum Graphity \cite{MarkHamma}.
 The target \textit{per se} is understanding the emergence of gravity through model building of discrete quantum 
systems. A similar approach is Analogue Models \cite{bvl}, although in the continuum and considering
Quantum Field Theory as the main tool. 
The motivation for the introduction of simplified models for Quantum Gravity is understanding the features of 
background independence in contexts in which not all the obstructions 
typical of gravity are present and, in particular, better understanding the phenomenon of emergence in background 
independent contexts. 

The first Quantum Graphity model was introduced in \cite{graphity1,konopka}. 
The basic motivation in \cite{graphity1} was to construct a Hamiltonian on a Hilbert space associated
to the degrees of freedom of a graph (a set of nodes $V$ with cardinality $\mathscr N$ and of links $\mathscr E$ of 
cardinality $\mathscr N (\mathscr N-1)/2$) such that the ground state of the Hamiltonian is a graph which has geometrical properties. 
The initial hope, in \cite{graphity1}, was that by fixing the parameters of the reduced model (the one with \textit{on/off} links), the ground state of the 
Hamiltonian would have been a graph of average degree $d$ and with geometrical properties 
similar to simplicial complexes\footnote{However the results obtained numerically in \cite{konopka}
suggested that the low-energy structure of the graph is string-like: the ground state is a one of a 
1-dimensional object. This result is compatible with the mean field
theory analysis performed in the reduced model\cite{ff}, in which the model was mapped to an Ising-type
Hamiltonian. This mapping allowed a 
straightforward use of the mean field theory techniques well known from the study of Ising-type models, 
after having identified the average degree of the graph as an order parameter. In \cite{conrady} a different Hamiltonian
based on graphs and close in spirit to \cite{graphity1} found, instead, a 2-dimensional complex in a low-energy phase. 
This last result gives some hope that a generic mechanism to obtain low energy $d$-dimensional simplicial complexes exists.

}. 

A second model, which we now describe,
was introduced in \cite{graphity2}. 
The main motivation for the introduction of \cite{graphity2} was the interpretational issue of the external bath (i.e.\ the temperature of the system):
how can it be interpreted in a closed system (i.e. the Universe)? The same problem arises in closed quantum systems, where one could ask why thermalization is such a general phenomenon despite the 
unitary time evolution on the Hilbert space. In general the solution to the problem is strongly dependent on the \textit{observability} of full Hilbert space, i.e.\ what part of it should be traced 
out in order to observe decoherence in the local observables. From this perspective, \cite{graphity2} is a 
simplified model in which these questions can be answered in a background independent context. While the graph degrees of freedom are the same as the model introduced in 
\cite{graphity1},
additional degrees of freedom associated to bosonic particles with a Bose--Hubbard (BH) interaction are present. The result is a dynamical graph with local interactions.


\section{Matter coupling: a graph-dynamical Bose--Hubbard model}
The model introduced in \cite{graphity2} and recently further studied in \cite{graphity3}, focuses
 on the study of interaction between matter (in the specific case, bosonic degrees of freedom on the vertices 
of the graph) and the graph. The energy terms for links and particles are of the form
\begin{equation}
\widehat H_0=\sum_i \mu\ \widehat a^\dag_i \widehat a_i + \sum_{ij} \nu\ \widehat b^\dag_{ij} \widehat b_{ij},
\end{equation}
with $\widehat a_i$, $\widehat b_{ij}$ are hard core bosonic ladder operators on the space of vertices and links respectively and $\mu$,$\nu$ coupling constants. There are two other terms, namely a BH interaction,
\begin{equation}
\widehat H_{BH}=-E \sum_{ij}  (\widehat a^\dag_i \widehat a_j+h.c.) \otimes \widehat P_{ij},
\end{equation}
where $\widehat P_{ij}$ is a projector on the \textit{on} links $ij$, and an interaction between the bosonic 
particles and the links,
\begin{equation}
\widehat H_{int}=-E \sum_{ij}  \widehat P^L_{ij} (\widehat a^\dag_i \widehat a^\dag_j \otimes \widehat b_{ij}
+ h.c.).
\end{equation}
$\widehat P^L_{ij}$ is a nonlocal projector which annihilates graph states unless the link $ij$ is not 
on a triangle. This keeps the dynamics local. \\
\textit{Thermalization.} Simulations\cite{graphity2} showed that, in the long time regime, the classical model evolves into random graphs. 
In the quantum case, instead, the damping of oscillations of graphs observables, e.g.\ the degree of the graph, indicates
that the model thermalizes to a metastable state. We studied the case of a 4-vertices graph with a link turned off and the 
others on. We observed dumping for the vertex degree observables. For the 
studied case, the asymptotic state turned out to be a mixed state of the graph with the link turned on and the 
link turned off.


\section{Graphs and curved geometry}

\begin{figure}[h]

\centering

\subfloat[The $\mathscr K_N$ configuration.]{
\includegraphics[scale=0.5]{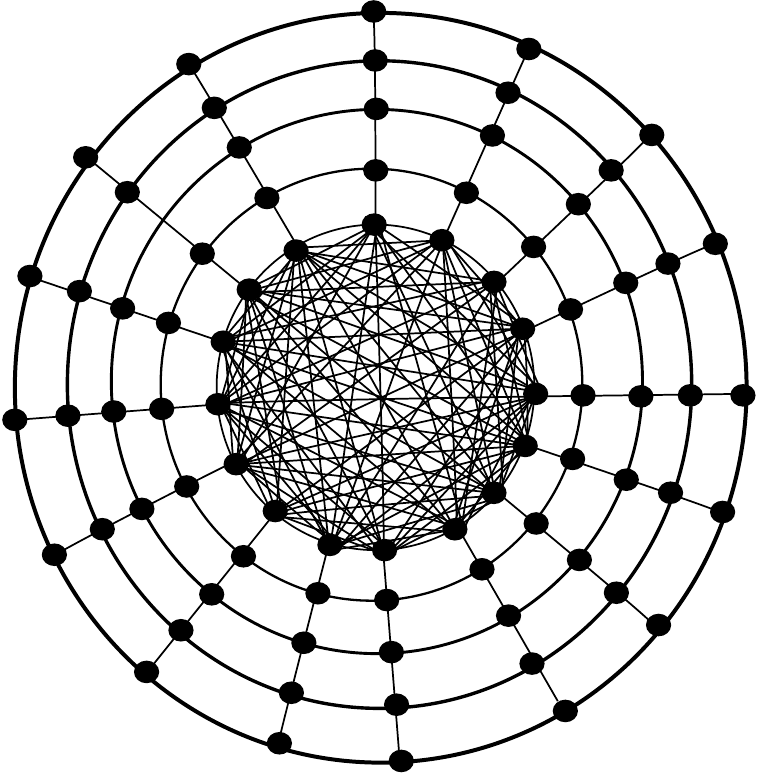}
\label{2dkn}
}
\subfloat[A non-trivial graph in which the coupling constants of the 1d-reduced model are site-dependent.]{
\includegraphics[scale=0.5]{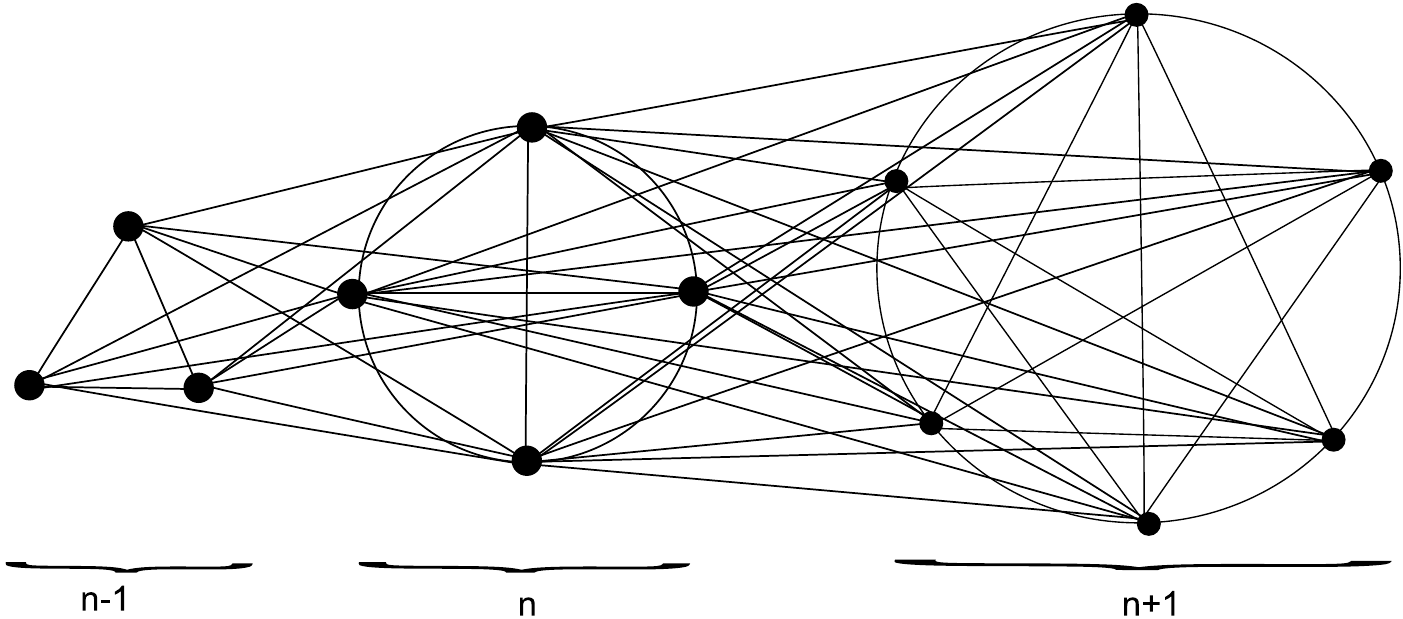}
\label{2dnt}
}
\caption{Two graphs on which the wavefunction can be symmetry-reduced.}
\label{2dfig}
\end{figure}

\textit{Trapped surfaces.} An interesting feature of this model, first noticed in \cite{graphity2}, is that regions of high connectivity 
are able to trap particles of low energy.  This was studied in detail in \cite{graphity3} on a fixed graph. The graph chosen is the one of 
Fig. \ref{2dfig}\subref{2dkn}, which has a rotational symmetry: single particle states can be labeled by two quantum numbers, the shell position
$n$ and the internal position, $\theta$, thus states of the form $|i,\theta\rangle$. In this case, the problem can be simplified by
noticing that the Hamiltonian is diagonal in blocks of constant angular momentum. We can thus introduce, in the 1-particle sector, the \emph{delocalized} states,
defined as:
\begin{equation}
 |n\rangle= \frac{1}{\sqrt{K_n}} \sum_{\theta=1}^{K_n} |n,\theta\rangle,
\end{equation}
where $K_n$ is the number of sites inside a shell $n$ (see Fig. \ref{2dfig}\subref{2dnt})
The full quantum evolution of a delocalized state can be \emph{fully} reduced to a 1-dimensional BH model if the initial state 
is delocalized, where the coefficients of the Hamiltonian are opportunely chosen. This allowed to prove that the ground state, 
for the graph in Fig. \ref{2dfig}\subref{2dkn}, is $|n=0\rangle$.
 
 In fact, this reduction showed that particles inside the highly connected central region of 
Fig. \ref{2dkn} feel a potential which is proportional to the degree. In particular, between particle states inside 
the central region and the outside regions there is an energy gap proportional the relative degree. 
This prevents low energy particles from escaping the trap, thus confirming the argument of \cite{graphity2}.\\
\textit{Emergent curved space.} Another interesting finding of \cite{graphity3} is that, in the reduced models, expectation values of number operators on the graph 
exhibit an emergent Lorentz symmetry. The reduced effective 1-dimensional BH Hamiltonian, 
for a generic rotationally invariant graph, takes the form,
\begin{equation}
\widehat H_{BH}=- \sum_{i} f_{i\ i-1}  (\widehat b^\dag_i \widehat b_{i-1}+h.c.),
\end{equation}
with $f_{ij}=d_{ij} E$ and $d_{ij}$ is the relative degree between the shell $i$ and the shell $j$, 
$\widehat b^\dag_i$,$\widehat b_i$ is a symmetry reduced ladder operators. The idea then is to study 
the following expectation values (e.v.) $\langle \widehat b^\dag_i \widehat b_i \rangle=\Psi_i(t) $. The interesting 
result is that these e.v. satisfy the following (closed) relation on the manifold of classical states:
\begin{align}
\frac{\hbar^2}{2}\partial_t^2 \Psi_n(t) =& f_{n-1,n}^2 \left(\Psi_{n+1}(t)+\Psi_{n-1}(t)-2 \Psi_{n}(t)\right) 
\nonumber  \\
&+\left(f_{n+1,n}^2-f_{n-1,n}^2\right)\left( \Psi_{n+1}(t)-\Psi_{n}(t)\right).
\label{eom}
\end{align}
This fully describes the time evolution of the probability density. In the continuum limit, eqn. (\ref{eom}) becomes
$$\Big[\partial_t^2 -\partial_x \Big(n(x) \partial_x\Big) \Big]\Psi(x,t)=0,$$
so it has a site-dependent speed of propagation for the density. This equation is everywhere locally Lorentz-invariant, with a local
speed of propagation given by $c_x=1/\sqrt{n(x)}$. The higher-dimensional version of this equation is known to be related to the Gordon background\cite{bvl}. In fact,
there is no obstruction to repeating the same procedure for graphs which can be foliated in more than one dimension. Thus, curved space is encoded in the connectivity of the 
graph (i.e.\ the local degree).

\section{Conclusions}
The emergence of gravity in a background independent contexts is a complete new research direction within the 
field of Analogue Models. In the case of graph-based models as ours it is not always obvious what are the right questions to ask. 
Indeed, we have shown in \cite{graphity3} that by focusing on a particular set of
graphs and by asking the right question, ``How does the particle probability density evolve?'', the phenomenon of emergence 
(in our case of a Lorentz symmetry)
can be identified. This happens because the graph we considered can be foliated, and then the states classified, according to this foliation. More recently, the same technique has been used to 
study the effects of disordered locality on the Lorentz symmetry, showing that a mass term appears \cite{disloc}.
We have identified graph configurations which, in a appropriate limit, can be considered as trapping surfaces. By considering delocalized states 
(i.e. symmetry-reduced wavefunctions),
we showed that high connectivity implies a high energy gap between particles being inside the region or outside. 

It is not clear, at this point, how general these models are and how many of the emergent phenomena of
Analogue Models can be reproduced. We believe that many interesting questions related to background independence (and emergent gravity \cite{egrev}) 
can be addressed in such simplified models.
The full connection between a graph and the low energy manifold associated to it is an open problem to address. 
It is tempting, given the large literature on 
Regge calculus, to associate a simplicial complex to the graph when possible. On the other hand, it may be more physical to require that only the scaling 
properties - for instance looking at the Heat Kernel of the Laplacian of the graph\cite{filk}- match those of manifolds of a specific dimension. 

{\footnotesize
\begin{center}
\textit{Aknowledgements}
\end{center}
We would like to thank A. Hamma, F. Markopoulou and A. Riera for several stimulating discussions and for the collaboration
in the main work that we summarized here. Also, we thank L. Sindoni for many fruitful conversations about emergent gravity.
Research at Perimeter Institute is supported by the Government of Canada through Industry 
Canada and by the Province of Ontario through the Ministry of Research \& Innovation. }
\begin{center}
\line(1,0){250}
\end{center}

\end{document}